# Formation of cellular/lamellar nanostructure in $Sm_2Co_{17}$-type binary and ternary Sm-Co-Zr magnets

Nikita Polin[1], Konstantin P. Skokov[2], Alex Aubert[2], Hongguo Zhang[2,3], Burçak Ekitli[2], Esmaeil Adabifiroozjaei[4], Leopoldo Molina-Luna[4], Oliver Gutfleisch[2], Baptiste Gault[1,5]

[1] Max-Planck Institute for Sustainable Materials, 0237, Düsseldorf 40237, Germany

[2] Functional Materials, Institute of Materials Science, Technical University of Darmstadt, Peter-Grünberg-Str. 16, 64287 Darmstadt, Germany

[3] College of Materials Science and Engineering, Key Laboratory of Advanced Functional Materials, Ministry of Education, Beijing University of Technology, Beijing 100124, China

[4] Advanced Electron Microscopy Division, Institute of Materials Science, Department of Materials and Geosciences, Technische Universität Darmstadt, Peter-Grünberg-strasse 2, Darmstadt 64287, Germany

[5] Department of Materials, Royal School of Mines, Imperial College, Prince Consort Road, London SW7 2BP, United Kingdom.

## Graphical abstract

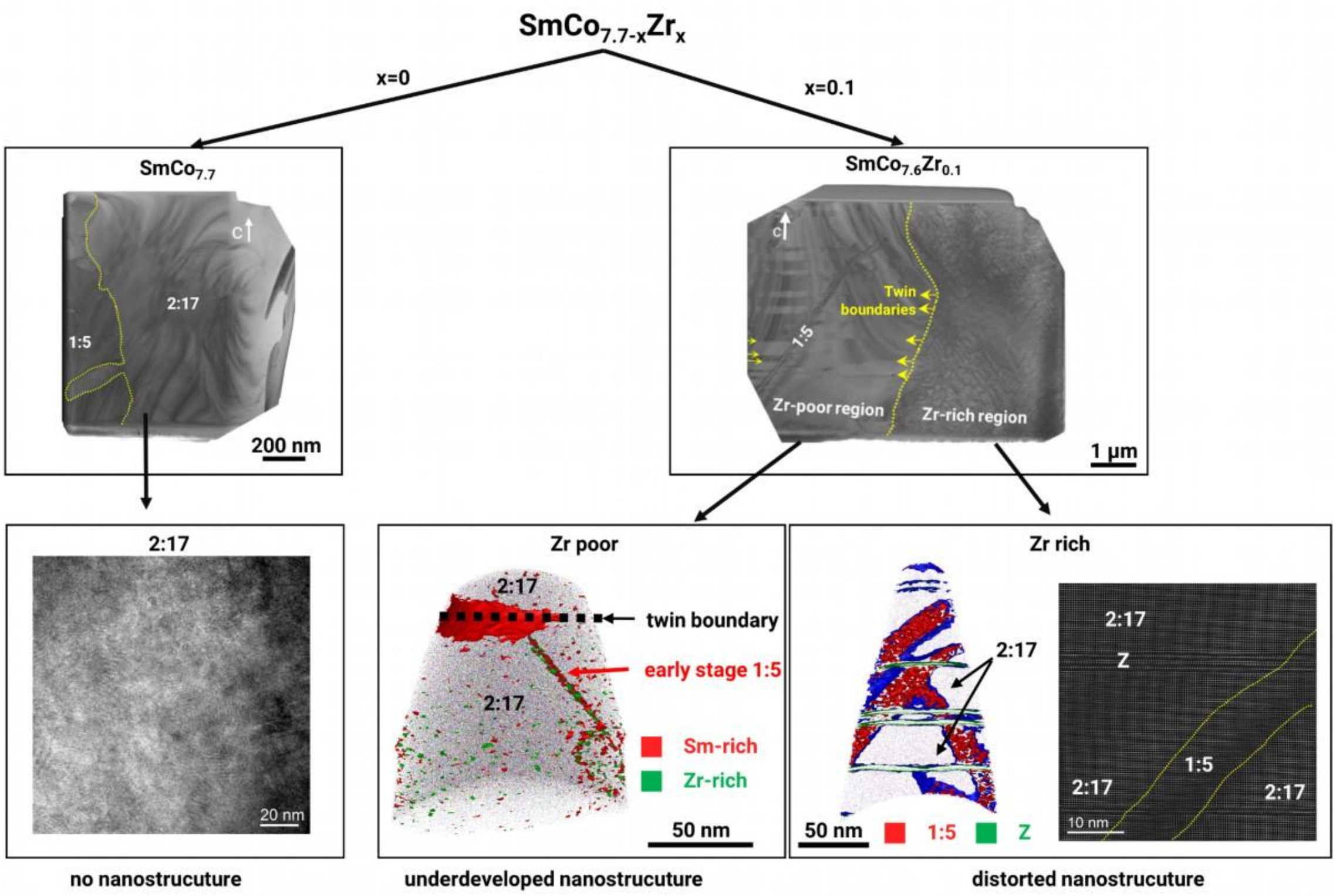

## Highlights

- Zr concentration above 1.4 at.% promotes cellular/lamellar nanostructure formation in ternary $SmCo_{7.6}Zr_{0.1}$
- In absence of Cu, small concentration gradients in the cellular nanostructure cause near zero coercivity in $SmCo_{7.6}Zr_{0.1}$
- 2:17 phase twins are suggested to facilitate the formation of 1:5 cell boundary phase

# Abstract

2:17 SmCo magnets with a quinary composition of $Sm(Co,Cu,Fe,Zr)_{7+\delta}$ are industrially-relevant hard magnets used in high temperature and corrosive environments. Their complex cellular/lamellar nanostructure, consisting of ordered 2:17 phase cells, 1:5 phase cell boundaries and Z-phase (Zr-rich) lamellae, is essential for their high coercivity. However, the system's complexity makes it challenging to determine the contribution of each element or microstructural feature to coercivity. To disentangle the microstructure-property relationships, we simplified the system to binary and ternary $SmCo_{7.7-x}Zr_x$ (with $x = 0$ and 0.1) magnets and conducted detailed micro- to atomic-scale analyses. Only Zr-containing magnets formed a cellular/lamellar nanostructure akin to industrial magnets, in Zr-rich regions with at least 1.4 at.% Zr, but without achieving high coercivity due to low elemental gradients in absence of Cu across cell boundaries. Data from Zr-poor areas of $SmCo_{7.6}Zr_{0.1}$ suggests that 2:17 phase twin boundaries facilitate cellular nanostructure formation by providing inhomogeneities for heterogeneous nucleation.

## Manuscript

The exceptional magnetic properties of 2:17-type Sm-Zr-Co-Cu-Fe hard magnets at elevated temperatures and in corrosive environments make them ideal for applications like sensors, electric vehicles, renewable energy and aeronautics [1–4]. Industrially sintered magnets have a quinary composition of $Sm(CoFeCuZr)_{7\pm\delta}$ and consist of ~ 100 μm diameter grains, textured along the c-axis of the $Sm_2Co_{17}$ phase. These grains exhibit a cellular/lamellar nanostructure with three main components [5,6]: the matrix $Sm_2(CoFeCuZr)_{17}$ (2:17) phase forms cells (~100 nm), separated by the $Sm(CoCuFeZr)_5$ (1:5) phase cell boundaries (~10 nm) and intersected by the Zr-rich (Z) phase lamellae (<10 nm) with stoichiometry close to $Sm(CoZrFeZr)_3$ perpendicular to the c-axis of $Sm_2Co_{17}$. This nanostructure forms during heat treatment, called precipitation hardening or aging, and is crucial for high coercivity [7,8].

The genesis of this cellular/lamellar nanostructure [9–12], as well as other factors influencing the magnets' coercivity, have been extensively studied for the quinary Sm-Zr-Co-Cu-Fe system, including composition [1,12–17], heat treatment parameters [6,12,18–22], stacking faults [23], oxidation [24], grain boundary hydrogen charging [25] and addition of non-magnetic particles [26,27]. However, the inherent complexity of this multicomponent, multiphase system makes isolating the role of each microstructural parameter on coercivity rather challenging. By reducing the alloy complexity, we can investigate step by step the formation of nanostructure and the development of coercivity. Here, we reduced the number of elements and defects by focusing on single grains (eliminating grain boundaries effects) of binary $SmCo_{7.7}$ and ternary $SmCo_{7.6}Zr_{0.1}$.

Few studies from the 2000s on precipitation hardened 2:17 type ternary Sm-Co-Zr alloys [2,15,16,28], found near-zero coercivity despite the presence of cellular/lamellar nanostructure, attributed to the lack of a large domain wall energy gradient across the 1:5 boundaries in absence of Cu [2,16]. However, the reported data lacked detailed structural and compositional data of the nanostructure needed to fully understand this absence of coercivity.

Here, using advanced microstructural characterization techniques including transmission electron microscopy (TEM) and atom probe tomography (APT), we provide the missing data on crystal structures, sizes and microchemistry of the nanostructure. We identified a 1.4 at.% Zr threshold for nanostructure formation and found further evidence for 2:17 twins enabling cellular structure formation [29]. Our findings enhance understanding of the cellular/lamellar nanostructure's genesis and enable a more rational design of quinary 2:17 Sm-Zr-Co-Cu-Fe alloys.

Elemental Sm, Co, and Zr (>99% purity) were alloyed via induction melting. The nominal compositions, $SmCo_{7.7}$ and $SmCo_{7.6}Zr_{0.1}$, are used throughout this study. To compensate for Sm loss during melting and annealing, 5 wt.% Sm was added. The molten alloys were crushed, wrapped in Mo-foil, sealed in a quartz tube under Ar, annealed at 1180°C for 1 hr for homogenization, and quenched in water. The alloys were aged at 850°C for 18 hrs, cooled to 400 °C at 0.5 K/min, held for 4 hrs, following protocols for quinary Sm-Zr-Co-Cu-Fe alloys [4], and subsequently crushed into single grains with sizes around 1-3 mm.

Magnetic properties were measured on ~1 mm wide single grains of $SmCo_{7.7-x}Zr_x$ the easy axis aligned along the external magnetic field direction using a room tem-perature vibrating sample magnetometer (VSM), without demagnetization field correction. For microstructure analysis, the ~3 mm wide single grains were aligned in a magnetic field with the *c*-axis out of plane, confirmed by electron-backscattered diffraction (EBSD), then embedded in a solid polymer, grinded, and polished. Microstructure was conducted by scanning electron microscopy (SEM, FEI Helios Nanolab 600i) using the backscatter electron mode (BSE), energy dispersive spectroscopy (EDX, FEI Helios Nanolab 600i) and electron backscatter diffraction (EBSD, Zeiss SIGMA 500). Magnetic domain imaging was performed by magneto-optical Kerr effect (MOKE evico magnetics GmbH) polarized light microscopy (Zeiss Axio Imager.D2m).

Electron transparent specimens for transmission electron microscopy (TEM) were cut out by Ga focused ion beam (FIB) in a SEM/FIB system (Dual-Beam Helios Nanolab 600i, FEI). Bright-field (BF) TEM imaging and selected-area electron diffraction (SAED) measurements were conducted in a conventional transmission electron microscope (JEOL JEM 2100F). High-resolution high angle annular dark field (HAADF) scanning TEM (STEM) imaging was carried out in an aberration-corrected system (JEOL JEM-ARM200F) operated at 200 kV.

For nanoscale composition, needle-shaped specimens were prepared by a Ga focused ion beam (FIB) (Dual-Beam Helios Nanolab 600i, FEI), following the procedure by Thompson et al. [30], using a low energy (5 keV) Ga beam for final milling to remove beam-damaged regions. These specimens were analysed using a CAMECA LEAP 5000 XS atom probe, operating at 40 K under ultra-high vacuum ($10^{-10}$ mbar), with pulsed UV laser (355 nm wavelength, 10 ps pulse duration, 40 pJ pulse energy, 200 kHz pulse rate) and a 1-2% detection rate. Datasets with at least 50 million ions were analyzed AP Suite v.6.3 by CAMECA.

**Figure 1** shows the magnetic hysteresis curves of $SmCo_{7.7}$ and $SmCo_{7.6}Zr_{0.1}$ single grains. Saturation occurs at external fields of ~0.5 T for $SmCo_{7.7}$ and ~1 T for $SmCo_{7.6}Zr_{0.1}$. Both samples exhibit near-zero coercivity, with $SmCo_{7.6}Zr_{0.1}$ showing slightly higher coercivity than $SmCo_{7.7}$, at $\mu_0 H_C$~20 mT and ~10 mT, respectively.

SEM-BSE mode images in **Figure 2a** reveal microstructural difference between the samples. $SmCo_{7.7}$ (**Figure 2a1**) displays a dark-contrast matrix phase and bright phase elongated inclusions, ~50 μm thick with an irregular shape. Matrix and inclusions are identified as $Sm_2Co_{17}$ (2:17) and $SmCo_5$ (1:5), respectively, as confirmed by EDX (**Table 1**). The higher than nominal Sm concentrations are likely due to the addition of Sm to compensate for a possible loss during processing. Smaller 2:17 islands are found within the 1:5 phase, while similarly, 1:5 phase islands are embedded in the 2:17 phase.

$SmCo_{7.6}Zr_{0.1}$ (**Figure 2a2**) exhibits dark contrast elongated inclusions (20 - 200 μm long, 10 - 20 μm wide), set in a brighter matrix. These inclusions are slightly richer in Sm (~1 at.%) and Zr (~0.5 at.%) and poorer in Co (~1.5 at.%), based on EDX (**Table 1**). EBSD (**Figure S1f**) confirms a uniform [0001] orientation of matrix and inclusions with grain size >500 μm. The intermediate grey contrast zones between inclusions and matrix (**Figure 2a2** bottom) result from their superposition as shown by TEM below. The phase assignment is ambiguous in the Sm-Co-Zr diagram [31], so matrix and inclusions are referred to as Zr-rich and Zr-poor regions.

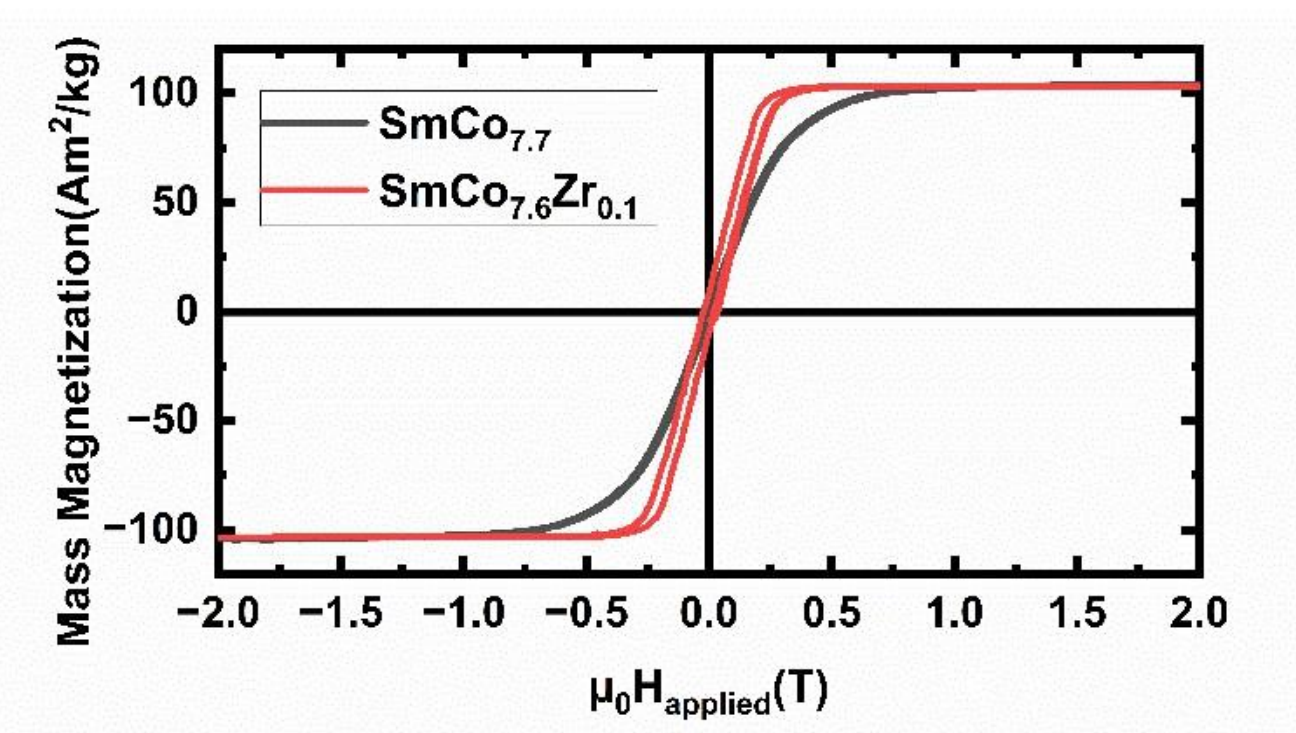

**Figure 1** Magnetic hysteresis curves of $SmCo_{7.7}$ and $SmCo_{7.6}Zr_{0.1}$, measured along the magnetically easy c-axis.

**Figure 2b** illustrates surface magnetic domain structures in Kerr microscopy images of the single grains, with nominal *c*-axis out of plane. In $SmCo_{7.7}$ (**Figure 2b1**), inclusions with coarser star-like domains resemble 1:5 phase, while finer domain regions match 2:17 phase. These structures are typical of high anisotropy magnets [32–34], indicating the single crystalline nature of both phases. The domain width of the 1:5 phase is larger compared to the 2:17 phase due to the higher anisotropy constant $K_1$ of the former [33,35], as the other two intrinsic magnetic properties, spontaneous magnetization $J_S$ and exchange stiffness constant $A$, have rather similar values.

In $SmCo_{7.6}Zr_{0.1}$ (**Figure 2b2**) finer domain inclusions correspond to Zr-poor regions, while coarser domain areas match the Zr-rich regions. The magnetic microstructure of the Zr-poor region resembles the 2:17 matrix phase in $SmCo_{7.7}$ (**Figure 2b1**), while for Zr-rich regions it resembles non-optimally cooled quinary $Sm(CoFeCuZr)_{7.2}$ alloys with near zero coercivity [17]. This suggests the presence of 2:17 phase in Zr-poor regions and of cellular nanostructure in the Zr-rich regions.

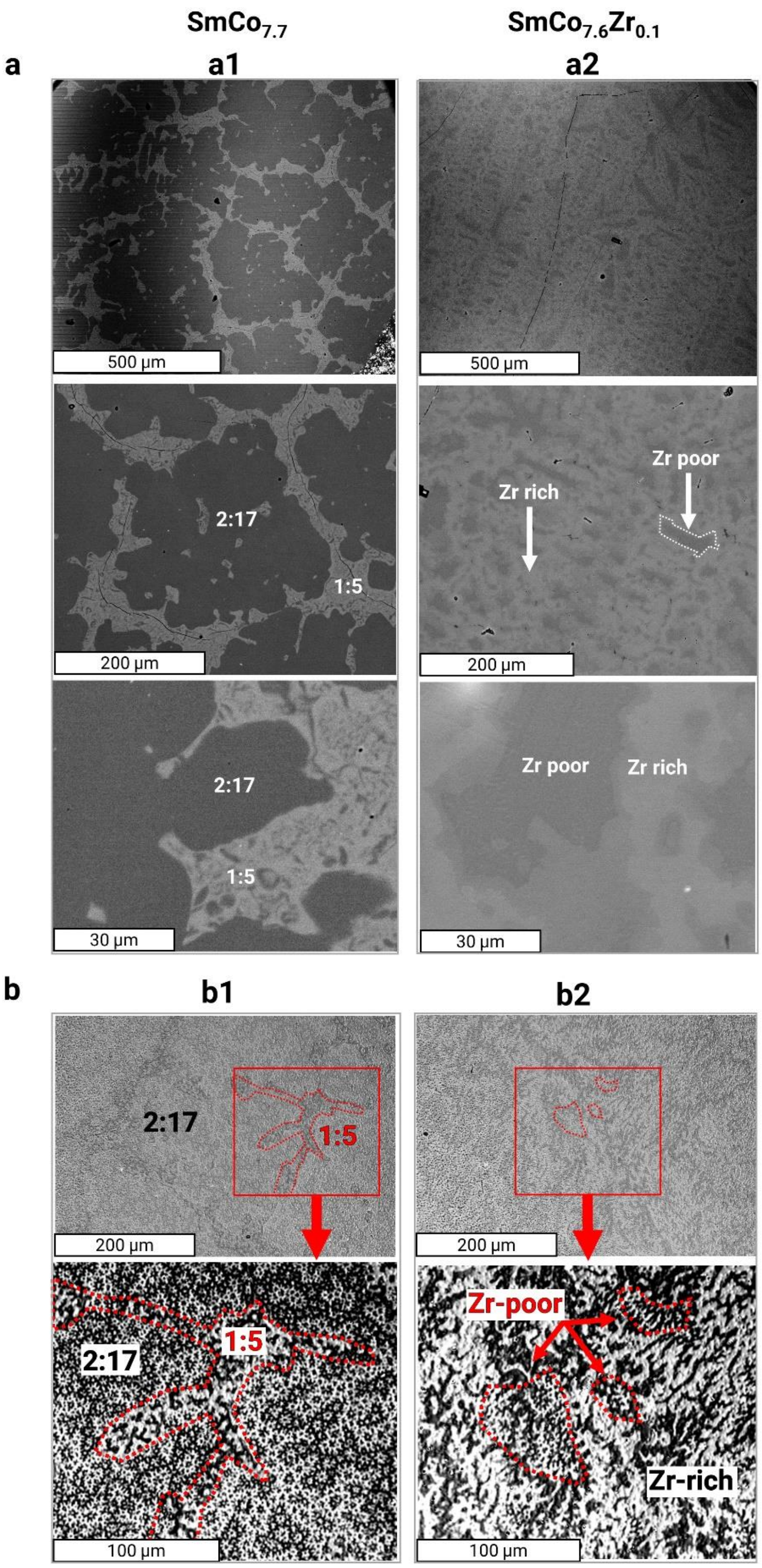

**Figure 2 a** Scanning Electron Microscopy in Backscatter Electron Mode (BSE SEM) and **b** magnetic domain structure by Kerr microscopy for $SmCo_{7.7}$ (left column, **a1, b1**) and $SmCo_{7.6}Zr_{0.1}$ (right column **a2,b2**). For Kerr and BSE, magnification increases from top to bottom. For all images the nominal *c*-axes of the single grains are oriented out of plane.

**Table 1** Compositions in respective regions of $SmCo_{7.7}$ and $SmCo_{7.6}Zr_{0.1}$ as determined by EDX and APT. For the composition column, Sm subscript was set to 1 for easier comparison

| Sample | Method | Region | Sm (at. %) | Co (at. %) | Zr (at. %) | Composition |
|---|---|---|---|---|---|---|
| $SmCo_{7.7}$ | EDX | 2:17 phase | 11.77 | 88.23 | - | $SmCo_{7.50}$ |
| | | 1:5 phase | 16.8 | 83.2 | - | $SmCo_{4.95}$ |
| $SmCo_{7.6}Zr_{0.1}$ | EDX | Zr poor | 12.21 | 86.91 | 0.87 | $SmCo_{7.12}Zr_{0.07}$ |
| | | Zr-rich | 13.41 | 84.85 | 1.73 | $SmCo_{6.33}Zr_{0.13}$ |
| | APT | Zr poor | 10.99 | 88.26 | 0.75 | $SmCo_{8.03}Zr_{0.07}$ |
| | | Zr-rich | 12.75 | 85.82 | 1.43 | $SmCo_{6.73}Zr_{0.11}$ |

In **Figure 3a1**, BF TEM images of $SmCo_{7.7}$ confirm phase separation into 1:5 and 2:17R phases, validated by electron diffraction (**Figure S2**). No cellular or lamellar nanostructure is evident in either the electron diffraction images or the HRTEM image of the 2:17 phase (**Figure 3a2**).

In $SmCo_{7.6}Zr_{0.1}$, the BF image (**Figure 3b1**) reveals distinct Zr-poor and Zr-rich regions**,** separated by a well-defined boundary without an intermediate zone. The Zr-poor region primarily consists of the 2:17R phase, enriched in 2:17 twins, as seen from abrupt contrast variations in the BF image and supported by electron diffraction (**Figure S2**), with an estimated area boundary density of $2.7\ \mu m^{-2}$ [36]. A single 100 nm thick 1:5 phase stripe is inclined at 35° to the c-axis. No cellular nanostructure was observed, aligning with Kerr microscopy results.

The Zr-rich region in $SmCo_{7.6}Zr_{0.1}$ exhibits a developed cellular and lamellar nanostructure, evident from BF images (**Figure 3b1** and **2b2**), HR-TEM images (**Figure 3c1**) and confirmed by electron diffraction (**Figure S2**). Compared to typical quinary Sm-Co-Cu-Fe-Zr alloys [1,7,11], three key distinctions are noted: First, the 2:17R phase cells are closer to parallelograms, unlike the typical diamond shape, with unequal lengths of 50 − 200 nm. Second, the Z-platelets are relatively thin and, third, not always continuous, with an average thickness to 1.6 nm based on FFT (reflection at $\frac{1}{4}$ (003)), smaller than the 5-10 nm typical in quinary alloys. Atomic resolution HAADF images (**Figure 3d1 and 2d2**), confirm the atomic structures of the phases, classifying the observed Z-phase as type II [37]. Additionally, a single 2:17 layer between two Z-platelets resembles the observations 1:5 phase layers between Z-platelets in quinary Sm-Co-Cu-Fe-Zr magnets [38].

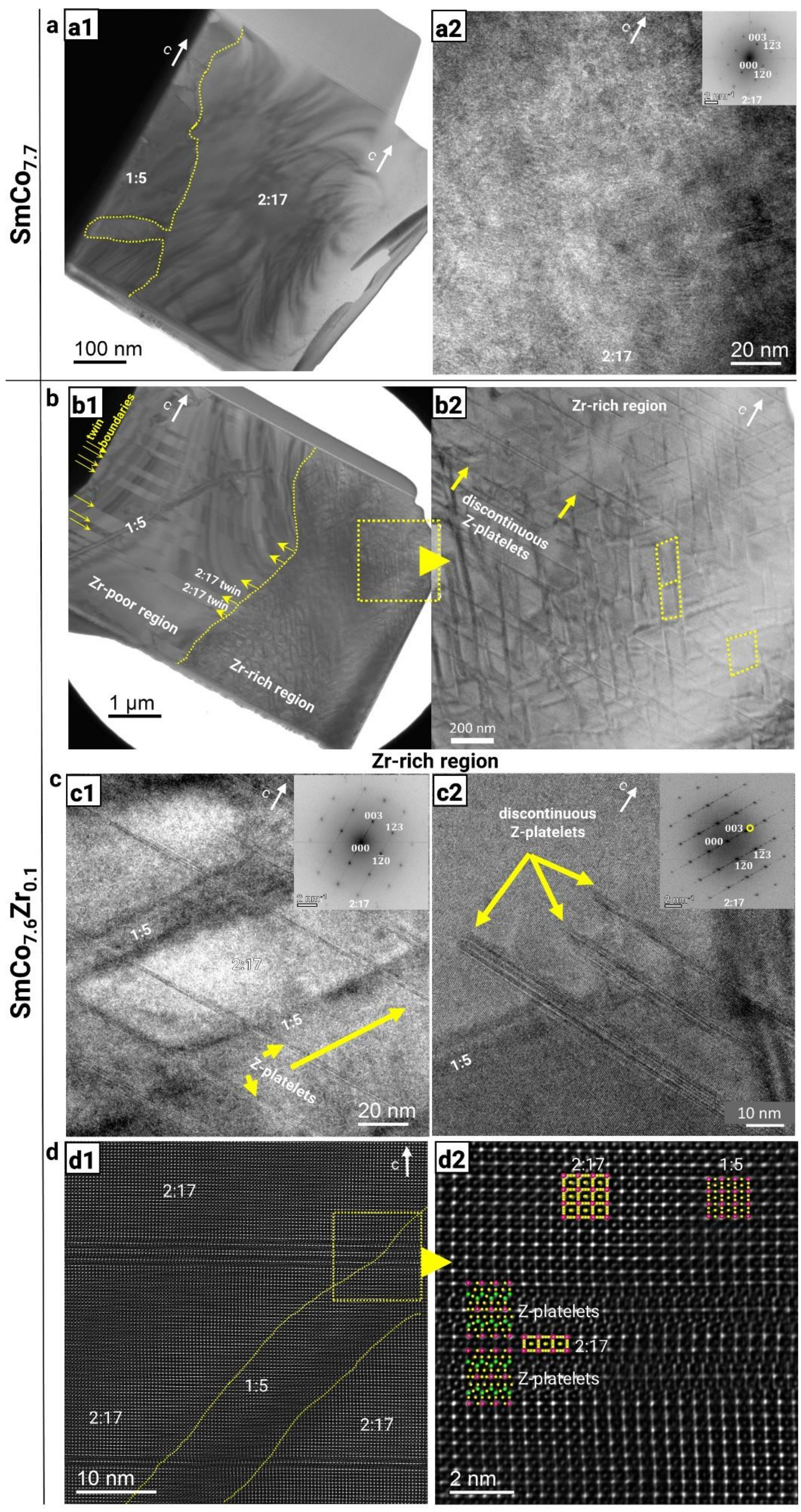

**Figure 3** TEM and STEM Characterization of $SmCo_{7.7}$ (**a**) and $SmCo_{7.6}Zr_{01}$ (**b-d**) with insets being FTTs of the respective image. **a** TEM images of $SmCo_{7.7}$: **a1** low magnification BF image with phase boundary between 1:5 and 2:17R phases in $SmCo_{7.7}$ marked in dashed yellow and **a2** HR-TEM of the 2:17 phase in $SmCo_{7.7}$ (b) BF TEM images of $SmCo_{7.6}Zr_{0.1}$ at **b1** low magnification illustrating the Zr-poor and Zr-rich regions, with boundary in dashed yellow and **b2** digitally magnified view on the Zr-rich region with parallelogram shaped cells marked dashed yellow. **c** High magnification TEM images of $SmCo_{7.6}Zr_{0.1}$ within the Zr-rich region: **c1** HR TEM, **c2** high magnification BF TEM. **d** Filtered HAADF STEM images of $SmCo_{7.6}Zr_{0.1}$ at **d1** low and **d2** high magnification with overlaid crystal structures of the corresponding phases.

**Figure 4a1** and **b1** present 3D APT reconstructions for $SmCo_{7.6}Zr_{0.1}$, showing Zr-rich and Zr-poor regions. The *c*-axis aligns with the APT specimen's main axis, confirmed by the Z-platelets' orientation and atom probe crystallographic analysis [39] (**Figure S5**). APT reveals that Zr-rich regions contain more Sm (~2 at.%) and Zr (~0.7 at.%) and less Co (-2.5 at.%) compared to Zr-poor regions, consistent with EDX data (**Table 1**), with corresponding mass spectra shown in **Figure S3**. Using isoconcentration surfaces, the data are segmented into Sm-rich (red) and Zr-rich (green) areas with thresholds reported in caption of **Figure 4**.

In the Zr-rich region of $SmCo_{7.6}Zr_{0.1}$ (**Figure 4a**), Sm-rich and Zr-rich areas correspond to 1:5 phase and Z-phase, the former having the highest solubility of Sm and the latter for Zr compared to competing phases [40,41]. The interfaces to the 2:17 phase are marked in blue. A 3D network of ~10 nm thick 1:5 phase cell boundaries surrounds parallelogram-shaped 2:17 cells, with ~5 nm thick Z-platelets intersecting perpendicular to the c-axis, as seen from the complete APT dataset (**Figure 4a1**) and a 10 nm-thick dataset slice (**Figure 4a2**). Although the cell size could not be estimated due to the limited field-of-view of APT, the observed nanostructure agrees with TEM.

Phase composition from representative 1D profiles (**Figure 4a3** and **3a4**) are: $Sm_1Co_{5.1}Zr_{0.1}$ for the 1:5-phase (arrow #1), $Sm_1Co_{10.9}Zr_{1.9}$ for the Z-phase (arrow #2) and $Sm_2Co_{16.1}Zr_{0.1}$ for the 2:17 phase. The 1:5 phase has ~0.5 at.% more Zr than 2:17 phase, indicating higher Zr solubility, likely due to preference of Zr to occupy the Sm-sublattice due to similar atomic radius [31]. Zr concentration and stoichiometry (Sm+Zr to Co) in the Z-phase, close to 1:4, resemble quinary Sm-Co-Fe-Cu-Zr magnets [38,42,43]. The elemental gradients across the 1:5-2:17 phase boundary (in $\mathrm{at.\% \cdot nm^{-1}}$ for Sm; Co; Zr : 2.3; 2.4; 0.3) are much smaller than in quinary Sm-Co-Fe-Cu-Zr magnets ($6 - 18\ \mathrm{at.\% \cdot nm^{-1}}$ for Cu) [38,43].

In the Zr-poor region of $SmCo_{7.6}Zr_{0.1}$, the 3D composition map and 3D point cloud (**Figure 4b1** and **b2**) reveal a mostly homogeneous elemental distribution, with composition $Sm_2Co_{16.06}Zr_{0.14}$ close to 2:17 stoichiometry with Sm-excess and 0.1 at.% dissolved Zr. However, distinct features were identified, with compositions taken from 1D profiles (**Figure 4b3** and **b4**): (1) A planar, mostly segregation free feature, identified as a twin boundary based on atom probe crystallography analysis [39] (**Figure S5**); (2) A Sm-Zr-rich planar feature adjacent to the twin boundary (arrow #3), inclined at ~40° to (006) planes and coherent with the 2-17 phase (**Figure S4**), with composition $Sm_1Co_{6.6}Zr_{0.1}$ indicating early-stage 1:5 phase with Co excess and a Sm deficit; (3) A Sm-rich feature (arrow #4) partially covering the twin boundary, with a composition $Sm_1Co_{1.48}Zr_{0.02}$ not assignable to any known phase from the phase diagram [44], indicative for a metastable phase, possibly formed due to local fluctuations of Sm, Co and Zr. These features were observed in one out of three APT measurements, with the others showing a homogenous distribution, consistent with twin density in TEM.

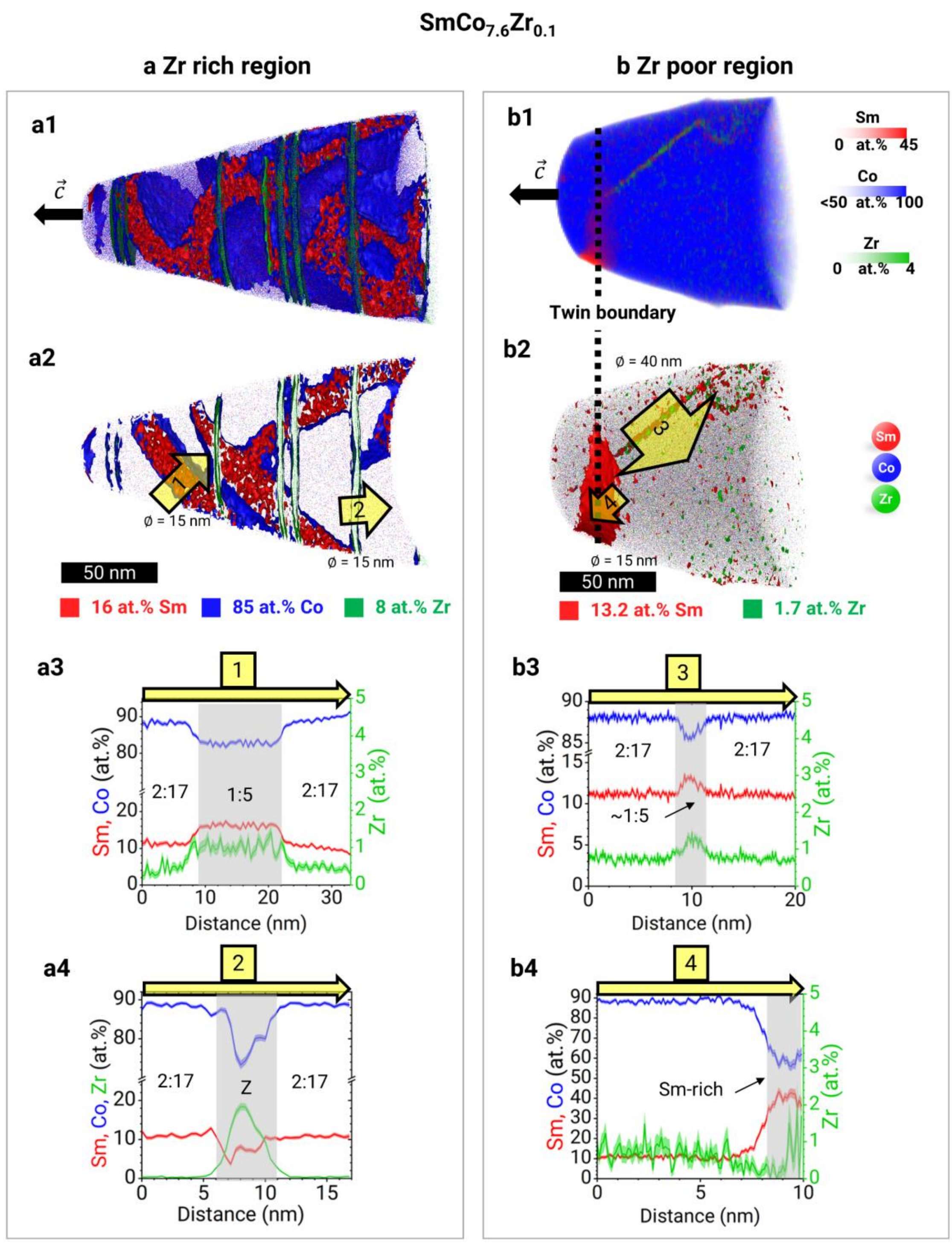

**Figure 4** 3D APT characterization of $SmCo_{7.6}Zr_{0.1}$ in the respective regions – **a** Zr-rich region and **b** Zr-poor region. **a1-a2** Complete dataset and 10-nm slice of the 3D APT reconstructions for Zr-rich region showing the geometrical distribution of the 1:5 phase (red, isoconcentration value 16 at.% Sm), the Z-phases (green, isoconcentration value 8 at.% Zr) and their interfaces to the 2:17 phase (blue, isoconcentration value 85 at.% Co). **b1-b2** 3D concentrations and 3D APT reconstruction for Zr-poor region reveling an early stage 1:5 phase and Sm-rich feature at the twin boundary using isosurfaces for Sm (red, isoconcentration value 13.2 at.% Sm) and Zr (green, isoconcentration value 1.7 at.% Zr). **a3,a4,b3,b4** representative 1D profiles across features observed in the 3D APT data as indicated by enumerated yellow arrows.

**Figure 5** summarizes our findings on the roles of (1) Zr and (2) twins on the cellular/lamellar nanostructure formation in $SmCo_{7.7-x}Zr_x$, and (3) the impact of chemical gradients on coercivity:

(1) In $SmCo_{7.7}$ (no Zr), only separated 1:5 and 2:17R SmCo phases form, without any nanostructure, consistent with literature [16,31]. During homogenization at 1180°C, the absence of Zr prevents the stabilization of a high-temperature phase, likely causing alloy decomposition before quenching [31]. Zr addition, however, enables the metastable 2:17H phase, ensuring alloy homogeneity before quenching [31].

In $SmCo_{7.6}Zr_{0.1}$ (with Zr), regions with over 1.4 at.% Zr (Zr-rich) form a cellular/lamellar nanostructure with distorted cells and discontinuous, thin Z-platelets, while insufficient Zr < 1.4 at.% leads to undesired Zr-poor regions without a developed cellular/lamellar nanostructure. To our knowledge, the 1.4 at.% Zr threshold for nanostructure formation in SmCo based alloys has not been previously reported. This threshold is met in previous reports on ternary $Sm(Co,Zr)_{7+\delta}$ alloys (2.5 at.% Zr in Refs. [2,45]) and in commercially used quinary $Sm(Co,Cu,Fe,Zr)_{7+\delta}$ magnets Cu and Fe containing (1.8 – 1.9 at.% Zr in Refs. [38,43]). These findings support the hypothesis that Zr-stabilized Z-platelets facilitate the diffusion for other elements and the formation of the cellular/lamellar structure [12]. Optimized Zr content and longer homogenization may eliminate Zr-poor regions and promote an optimal nanostructure throughout the sample.

(2) The Zr-poor regions are rich in 2:17 twins, which were suggested to aid lamellar [1,4,46] and cellular [29] nanostructure formation during the heat treatment process for quinary Sm-Zr-Co-Cu-Fe magnets. Twins are common in magnetic systems [47–49] and may be induced by quenching stresses, similar to carbon steels. As SmCo magnets are low-crystal-symmetry materials (hexagonal or rhombohedral) and hence inherently brittle, twinning is expected to be the favoured mechanism to accommodate thermal stresses, occurring during quenching, as the dislocation mobility and hence slipping are limited [51,52]. Additionally, twins might form to reduce the lattice strain energy caused by the coexistence of the 1:5 and 2:17 phases [53]. Early stage cellular nanostructure formation, as suggested by Wu et al. [29], is mitigated by 2:17R microtwin migration, creating antiphase boundaries, serving as nucleation centers for the 1:5 cell boundary phase as well as for Cu/Fe elemental segregation in this phase. The presence of the 1:5 phase in Zr-poor regions, observed at low fractions and thicknesses of 4-100 nm, indicates such an early stage cellular nanostructure, similar to Sm-Zr-Co-Cu-Fe magnets at the early stage of heat treatment [18,19]. This underdeveloped structure likely results from the local Zr deficit, with the 4 nm-thick early stage 1:5 phase, as revealed by APT, potentially representing an antiphase boundary. This finding further evidences 2:17R twins to aid cellular nanostructure formation.

(3) Both alloys exhibit near-zero coercivity: in $SmCo_{7.7}$ due to absence of a developed nanostructure, in $SmCo_{7.6}Zr_{0.1}$ despite a cellular/lamellar nanostructure present in >50% of the sample's volume. Similar findings were attributed to small concentration gradients across the cellular phase in Sm-Co-Fe-Ni-Zr alloys [12] and non-aged Sm-Zr-Co-Cu-Fe alloys [16,34]. In $SmCo_{7.6}Zr_{0.1}$, small gradients in Sm, Co and Zr across the 1:5/2:17R phase boundaries, and the absence of gradients within the 1:5 phase itself, results in minimal domain wall energy gradients and near-zero coercivity, as was proposed in previous studies for these alloys [2,16]. APT quantitatively confirmed these small gradients for $SmCo_{7.7-x}Zr_x$, highlighting that nanostructure alone does not guarantee high coercivity in SmCo magnets.

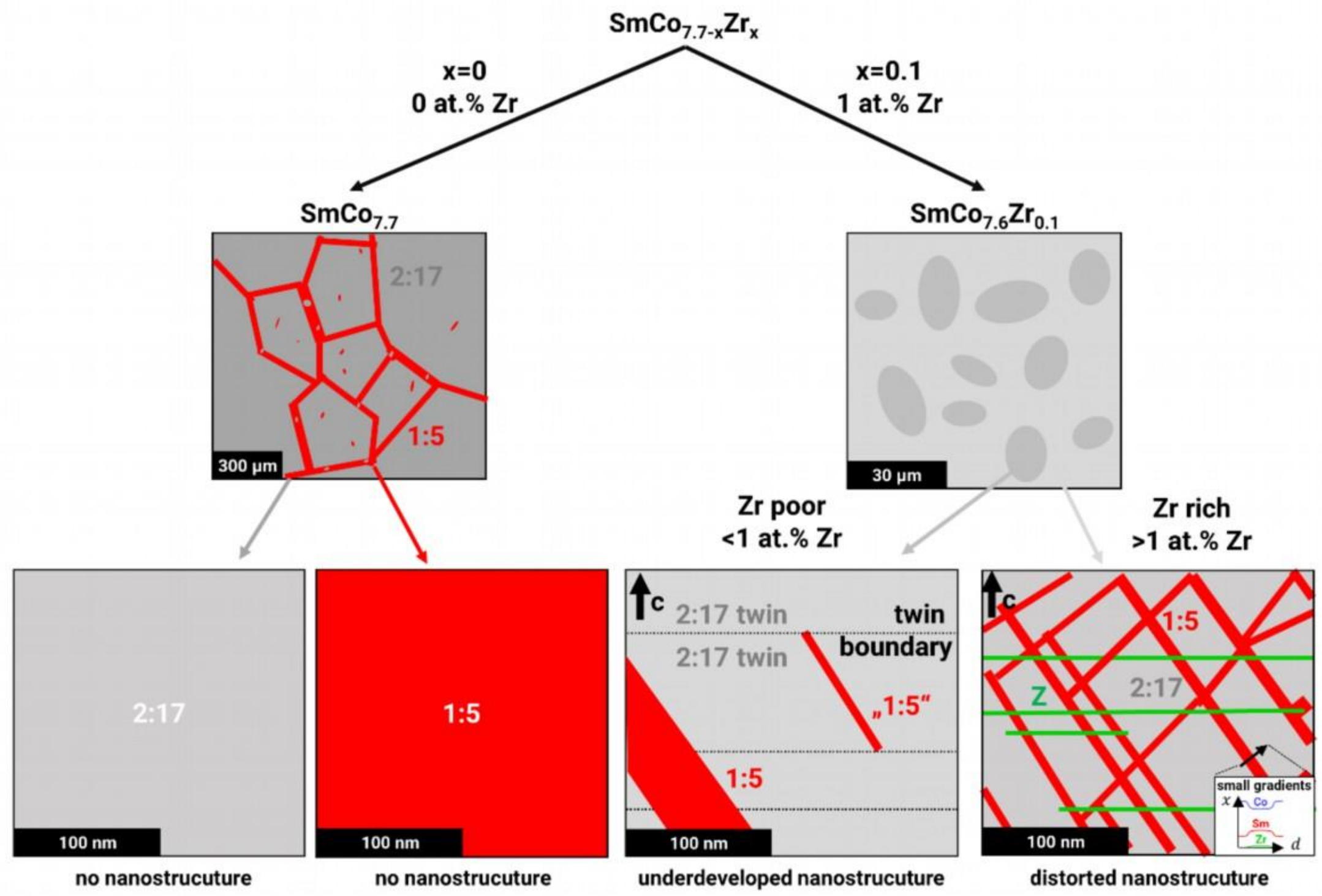

**Figure 5** Schematic representation of microstructure (top) and nanostructure (bottom) of $SmCo_{7.7-x}Zr_x$ alloys with $x = 0$ and 0.1: without Zr addition ($SmCo_{7.7}$ , left), the alloy separates into two pure phases being 1:5 and 2:17 phases without any cellular/lamellar nanostructure. With addition of Zr ($SmCo_{7.6}Zr_{0.1}$, right) into the alloy, two regions form with slightly different Zr and Sm content form: the Zr-poor region forms a strongly underdeveloped cellular nanostructure consisting of twinned 2:17R phase and rarely found 1:5 phase, while a Zr-rich region forms a distorted cellular/lamellar nanostructure consisting of parallelogram shaped cells of 2:17R phase, 1:5 phase cell boundaries and discontinuous, thin Z-platelets with low gradients across the 1:5/2:17 phase boundary. Zr content above 1.4 at. % is needed for nanostructure formation.

In conclusion, magnetic properties and microstructure of aged $SmCo_{7.7}$ and $SmCo_{7.6}Zr_{0.1}$ single grains were analyzed from meso to nanoscale. Without Zr, the necessary cellular/lamellar nanostructure fails to develop, leading to mesoscale phase separation. Adding Zr above 1.4 at.% enables the desired nanostructure, akin to quinary Sm-Zr-Co-Cu-Fe magnets. Insufficient Zr in $SmCo_{7.6}Zr_{0.1}$ results in distorted cells and thin, discontinuous lamellae in the Zr/Sm-rich regions and creates Zr/Sm-poor regions, characterized by 2:17 phase enriched in twins, which likely contribute to nucleation of minor 1:5 phases.

Near-zero coercivity in $SmCo_{7.7}$ is due to absence of a developed nanostructure, while for $SmCo_{7.6}Zr_{0.1}$, it is due to minimal elemental gradients across the cellular nanostructure, resulting in low domain energy gradients. The study emphasizes the significant role of Zr and 2:17 twins in promoting cellular/lamellar nanostructure formation, while also underscoring the importance of nanoscale chemical gradients to attain high coercivity in SmCo magnets.

## Acknowledgements

We thank Uwe Tezins, Christian Broß, and Andreas Sturm for their support to the FIB & APT facilities at MPIE. NP is grateful for the financial support from International Max Planck Research School for Sustainable Metallurgy (IMPRS-SusMet). We thank the Collaborative Research Centre/Transregio 270 HoMMage and all project partners for the fruitful discussions and scientific input. This work was supported by the Deutsche Forschungsgemeinschaft (DFG, German Research Foundation), Project ID No. 405553726, TRR 270.

## Declaration of Competing Interest

The authors declare that they have no known competing financial interests or personal relationships that could have appeared to influence the work reported in this paper.

# Supplementary material:

# Formation of cellular/lamellar nanostructure in $Sm_2Co_{17}$-type binary and ternary Sm-Co-Zr magnets

Nikita Polin[1], Konstantin P. Skokov[2], Alex Aubert[2], Hongguo Zhang[2,3], Burçak Ekitli[2], Esmaeil Adabifiroozjaei[4], Leopoldo Molina-Luna[4], Oliver Gutfleisch[2], Baptiste Gault[1,5]

[1] Max-Planck Institute for Sustainable Materials, 0237, Düsseldorf 40237, Germany

[2] Functional Materials, Institute of Materials Science, Technical University of Darmstadt, Peter-Grünberg-Str. 16, 64287 Darmstadt, Germany

[3] College of Materials Science and Engineering, Key Laboratory of Advanced Functional Materials, Ministry of Education, Beijing University of Technology, Beijing 100124, China

[4] Advanced Electron Microscopy Division, Institute of Materials Science, Department of Materials and Geosciences, Technische Universität Darmstadt, Peter-Grünberstrasse 2, Darmstadt 64287, Germany

[5] Department of Materials, Royal School of Mines, Imperial College, Prince Consort Road, London SW7 2BP, United Kingdom.

# A. Microstructure and nanostructure characterization

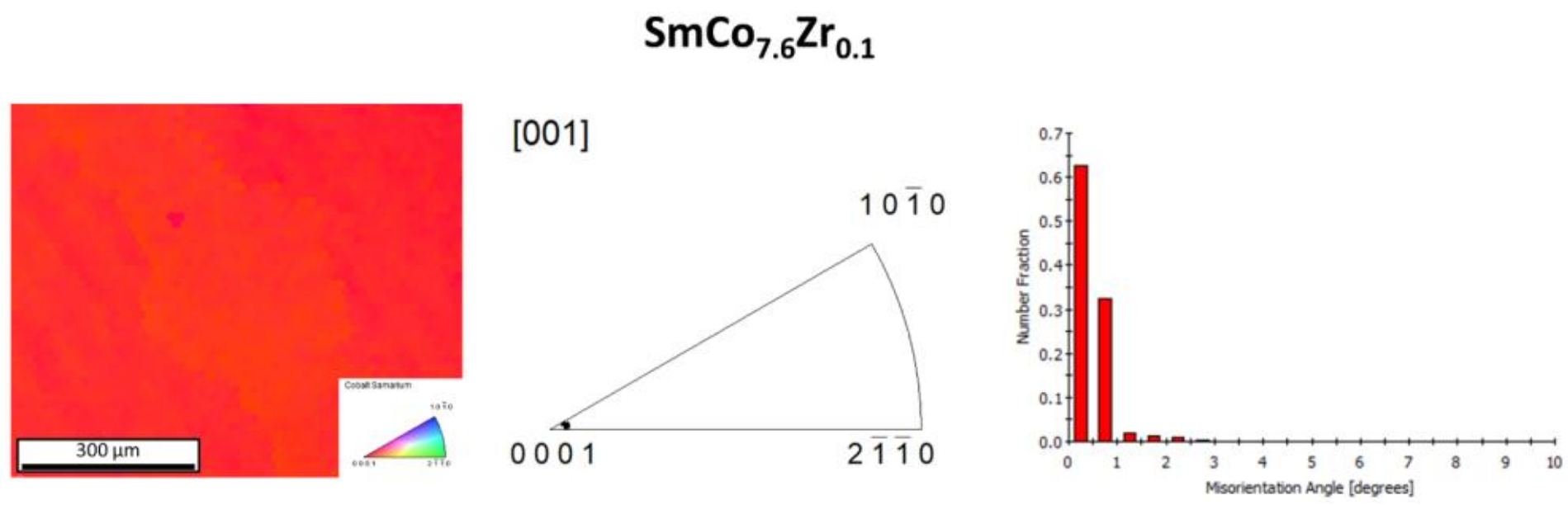

**Figure S1** Surface orientation characterization of $SmCo_{7.6}Zr_{0.1}$ by EBSD assuming of $Sm_2Co_{17}$ phase: [001] inverse pole figure map (left), inverse pole figure (middle), misorientation angle distribution (right). The sample has a uniform [001] orientation with misorientaion angles < 1°.

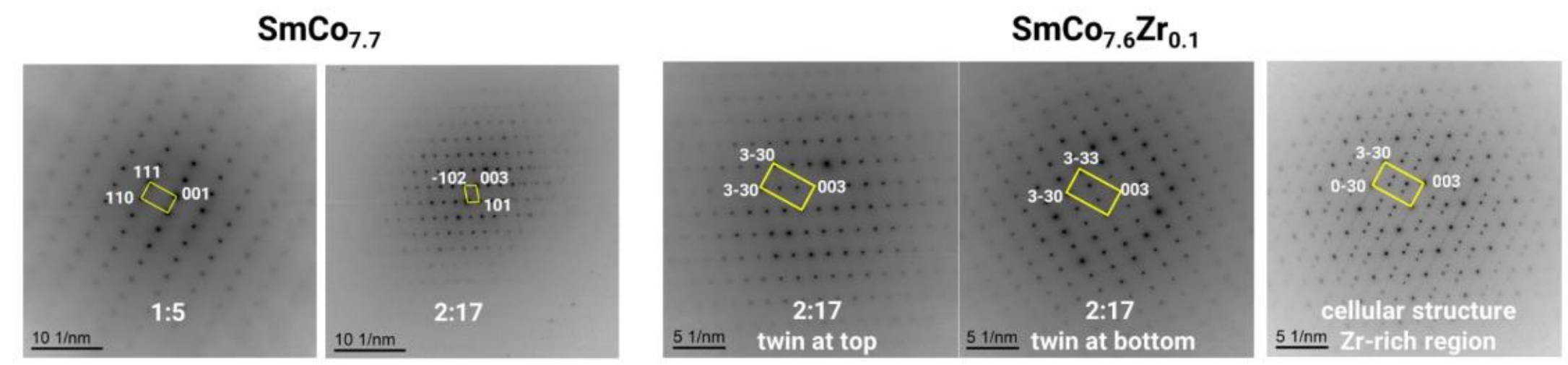

**Figure S2** Selected area electron diffraction (SAED) data extracted from regions of $SmCo_{7.7}$ (left two images) and $SmCo_{7.6}Zr_{0.1}$ (right three images) indicated in **Figure 3.**

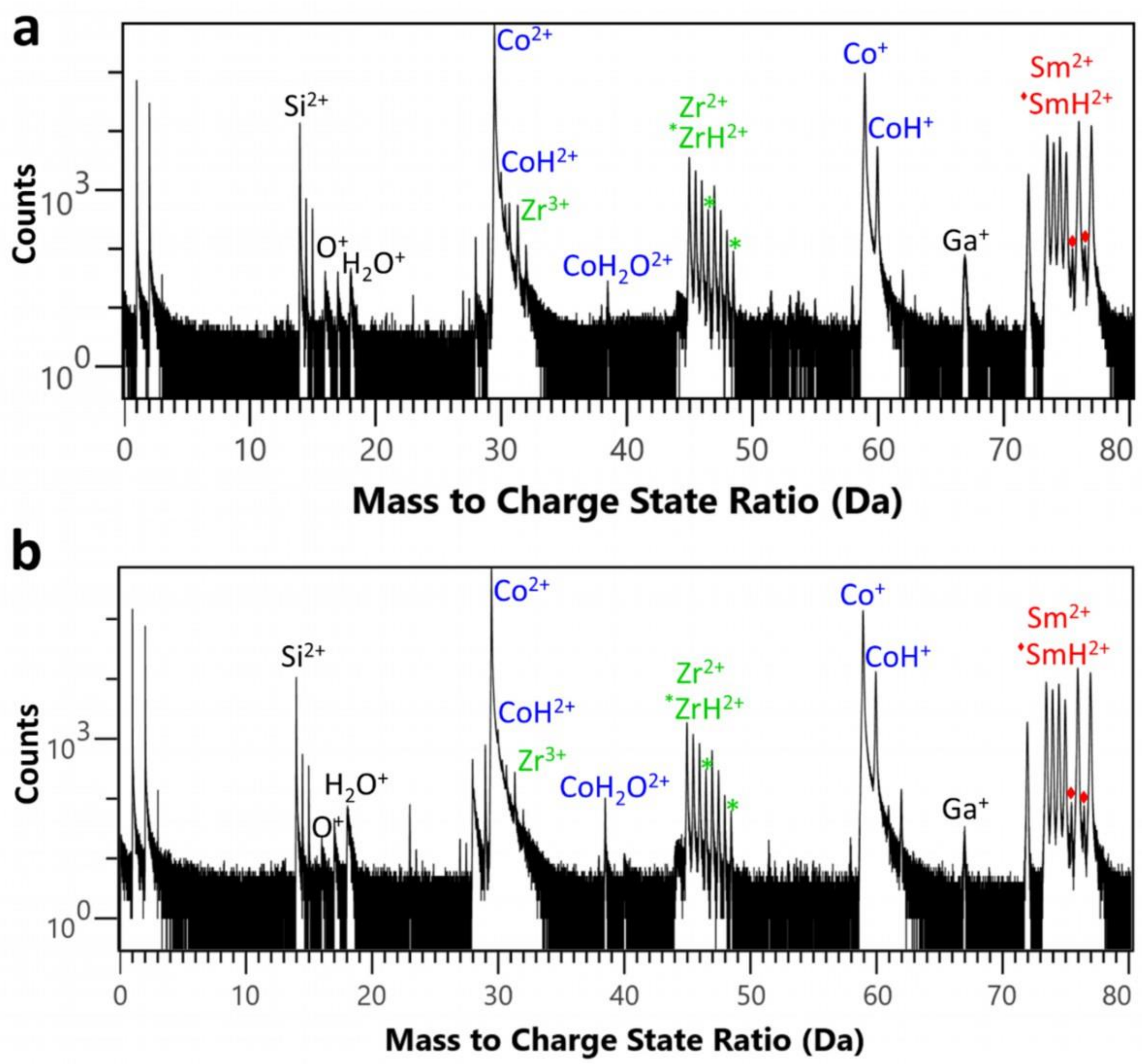

**Figure S3** APT mass spectra for **a** Zr-rich region and **b** Zr-poor region of of $SmCo_{7.6}Zr_{0.1}$.

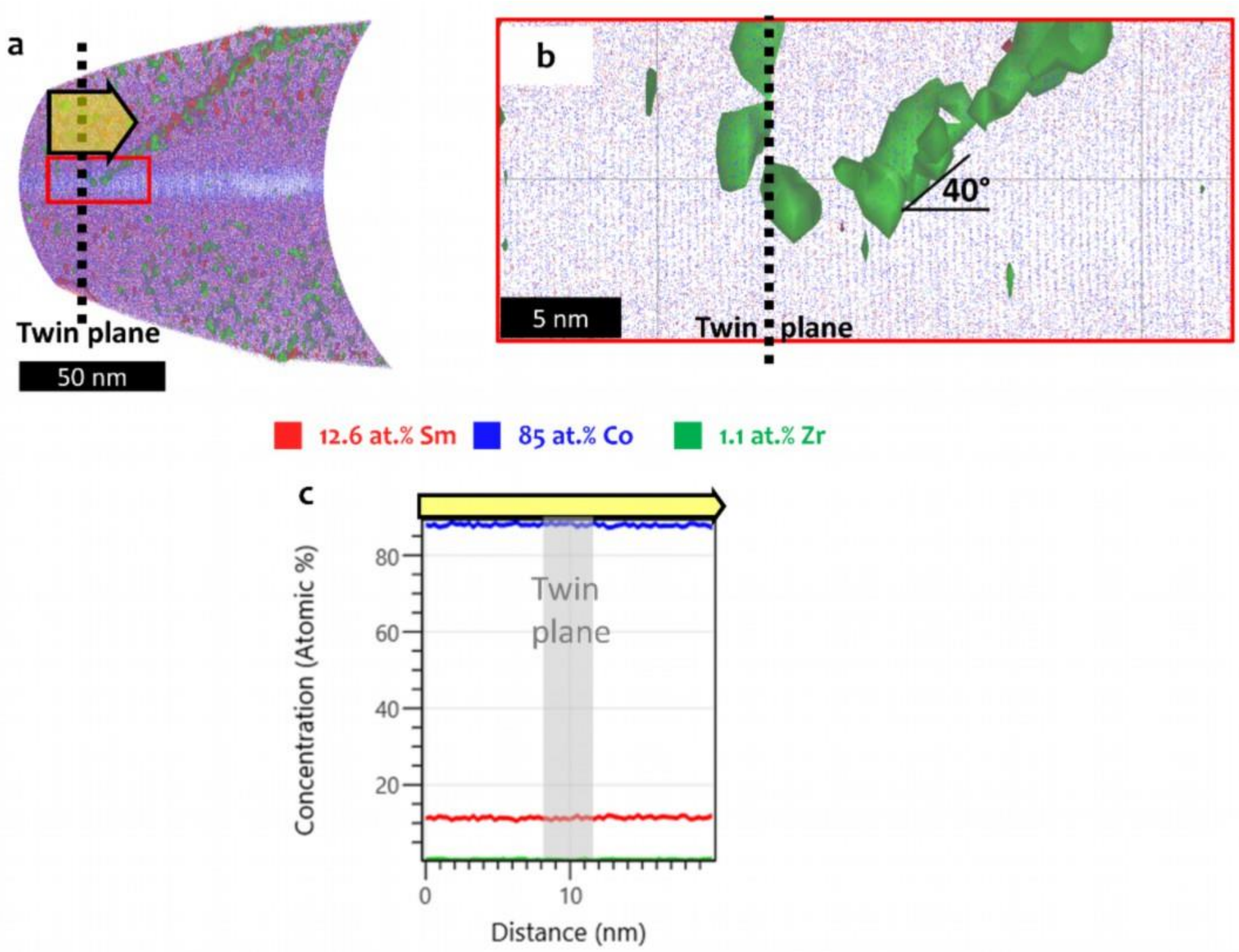

**Figure S4** Coherent precipitate close to early stage 1:5 phase at the twin boundary, in the 3D APT data in the Zr-poor region in the $SmCo_{7.5}Zr_{0.1}$ sample**. a** 10 nm scice of the 3D reconstruction of APT data shown in **Figure 4**b2. **b** Atomic planes in z-direction intersect the precipitate close to 1:5 phase in a magnified region of interest as marked by red rectangle in **a. c** 1D profile (cylinder 20nm length and diameter) across the twin plane (yellow arrow in **a**) which exhibits no elemental segregation away from the 1:5-like phase and the the Sm-rich feature The c-axis remains preserved at the twin plain as expected.

## B. Crystallographic analysis

For the crystallographic analysis, the following steps were followed:

1. Extract atomic hit map showing “poles” for the specific region of the APT dataset (i.e. top or bottom twin)
2. Overlay both hit maps with stereographic projection of the 2-17 phase with c-axis out of plane, such that hit maps and stereographic overlap
3. Identify the poles (h k I l) in the hit map and in the corresponding 3D data using the Co isovalues
4. Calibrate the reconstruction of the specific region such, that mean distance corresponds to the distance of the main pole or multiples of it: in this case [0006]
5. Measure the distances of identified poles (h k I l) in the 3D data from ROIs close to the poles
6. Compare the measured and expected distances of identified poles (h k I l): if the values equal, then step the poles were correctly identified. If not, repeat the procedure starting from step 2.

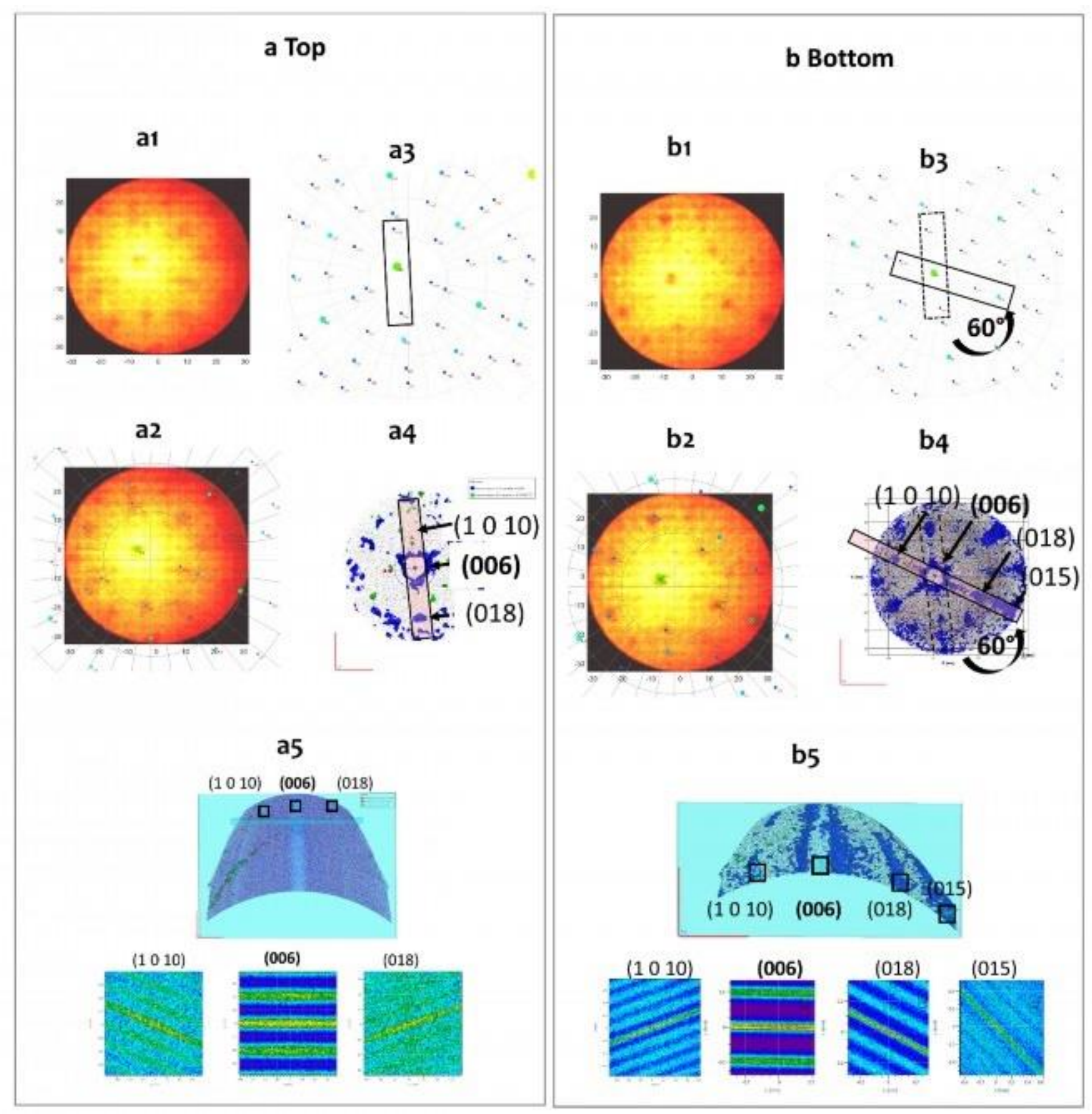

**Figure S5** Crystallographic analysis of the of the APT data for **a** Top part and **b** Bottom part for the Zr-poor region $SmCo_{7.5}Zr_{0.1}$ dataset: Top and bottom atomic hit maps (a1, b1), stereographic projections (a2,b2), overlay of the former (a3,b3) and slice though the 3D APT data displaying the poles observed in (a1, b1). (a5,b5) shows spatial density maps in xz-direction (xz-SDMs) extracted from regions of interest close to the poles determined before. Note that the projection of the bottom part is rotated by 60° in respect to the top part as shown by solid and dashed rectangles in b2 and b4 indicating a presence of a twin boundary.

**Table S1** Crystallographic analysis of the Zr-poor region $SmCo_{7.6}Zr_{0.1}$ APT data in the **Figure S5.**

| **Miller indices (hkl) of the poles** | **expected distances (nm)** | **measured distances, top region (nm)** | **measured distances, bottom region (nm)** |
|---|---:|---:|---:|
| (006) | 0.2 | 0.2 | 0.2 |
| (018) | 0.15 | 0.14 | 0.14 |
| (015) | 0.23 | - | 0.23 |
| (1 0 10) | 0.12 | 0.114 | 0.12 |